        \documentstyle[12pt]{article}
	\makeatletter
        \@addtoreset{equation}{section}
        \makeatother
        
	\newcommand\beq{\begin{equation}}
\newcommand\eeq{\end{equation}}
\newcommand\beqa{\begin{eqnarray}}
\newcommand\eeqa{\end{eqnarray}}
\newcommand\F{F^{\lambda\mu\nu\rho}}
\newcommand\f{F_{\lambda\mu\nu\rho}}
\newcommand\D{{\cal D}}

\def\a{A_{\mu\nu\rho}}
\def\A{A^{\mu\nu\rho}}
\def\X{x^{\nu\rho}}
\def\x{x_{\nu\rho}}

\def\veff{V^{\rm eff}(\phi)}

\newcommand\vcw{V_{CW}(\phi)}

\newcommand\disp{\displaystyle}

\begin{document}
	\begin{flushright}
	UTS--DFT--95--10
	\end{flushright}
        \begin{center}
        {\Large\bf Membrane Vacuum as a Type II Superconductor
	\footnote{This article is dedicated to the memory of Hiroomi
	Umezawa.}}\\
	\medskip
	\medskip
	{\large S.Ansoldi\footnote{E-mail address:
	ansoldi@vstst0.ts.infn.it}}\\
	{ Dipartimento di Fisica Teorica dell'Universit\`a,\\
	Strada Costiera 11, 34014-Trieste, Italy}\\
	\smallskip
	{\large A.Aurilia\footnote{E-mail address:
	aaurilia@csupomona.edu}}\\
	{ Department of Physics, California State Polytechnic
	University\\
	Pomona, CA 91768, USA}\\
	\smallskip
	{\large E.Spallucci\footnote{E-mail address:
	spallucci@vstst0.ts.infn.it}}\\
	{ Dipartimento di Fisica Teorica dell'Universit\`a,\\
	Istituto Nazionale di Fisica Nucleare, Sezione di Trieste,\\
	Strada Costiera 11, 34014-Trieste, Italy}

	 \thispagestyle{empty}
        \medskip

        \begin{abstract}
	We study a functional field theory of membranes coupled to a
	rank--three tensor gauge potential.
	We show that gauge field radiative corrections lead
	to membrane condensation which turns the gauge field into a
	{\it massive spin--0 field}. This is the Coleman--Weinberg
	mechanism for {\it  membranes}. An analogy is also drawn with
	a type--II superconductor. The ground state of the system consists
	of a two--phase medium in which the superconducting background
	condensate is ~``pierced''~ by four dimensional domains,
	or ~``bags''~, of non superconducting vacuum.
	Bags are bounded by membranes whose physical thickness is of
	the order of the inverse mass acquired by the gauge field.

        \end{abstract}
        \end{center}

	\section{Introduction}

	Of the many aspects of field theory explored by Umezawa during
his lifelong research activity, none seems more central and more far
reaching than the notion of ~``boson condensation''~ as a tool to
induce structure in the ground state of a physical system.
{\it Boson condensation may lead to the formation of extended
objects} \cite{wad}. This idea permeates Umezawa's work in the last
twenty five years and has inspired numerous original applications in
such diverse fields as condensed matter physics, gauge models of
particle physics and biology \cite{ume}.\\
In retrospect, recognizing the influence of Umezawa's ideas on our
own work, we have decided to investigate some new aspects of our
current research on the theory of extended objects against the
conceptual backdrop of the boson condensation approach. Even though
our discussion is applicable to a generic p--brane embedded in a
spacetime of arbitrary dimensions, the specific objects that we wish
to consider presently are {\it relativistic bubbles} (2--branes in
current terminology), because of their historic role in the
development of QCD via the formulation of the so called ~``bag
models''~ of hadrons and because of their increasingly important
role in modern cosmology. In either case, one has to deal with a
multiphase ground state characterized by the formation of domain
walls separating regions of spacetime with different values of the
vacuum energy density. Then, the question that we address in this
paper is the search of a mechanism capable of inducing such a
structure over the spacetime continuum. One possible answer, we
contend, involves the process of boson condensation, and we are
fairly confident that Umezawa would agree.
We are not equally confident, however, that he would endorse our overall
strategy without some qualifications. In fact, before plunging into a
technical discussion of our work, it seems appropriate to recall the
key conceptual steps of Umezawa's work for the sake of comparison
with our own approach.\\
{}From Umezawa's vantage point, spatially extended objects,
relativistic or not, arise as special solutions of {\it local
quantum field theories} through the process of boson condensation.
Some such solutions may have topological singularities, in the sense
that the curl of the gradient of the boson condensation function is
not necessarily zero. Once formed, extended objects may influence
the original quantum system. This ~`` back reaction''~ may be
accounted for by a self--consistent potential attributed to the
extended object.\\
The physical paradigm which reflects in full the above logical
sequence is a type--II superconductor. In this system, an external
magnetic field is squeezed into thin flux tubes by the vacuum
pressure of the Cooper pairs condensate. It is this picture that we
wish to extend to the case of relativistic bubbles minimally coupled
to an antisymmetric tensor gauge potential $\a(x)$. More
specifically, the purpose of this paper is twofold: first, we wish
to show how membrane condensation takes place inducing a two--phase
structure in spacetime; second, we wish to show that membrane
condensation can be driven by the quantum corrections of the gauge
field $\a(x)$, in analogy to the Coleman--Weinberg mechanism \cite{cw}.\\
All of the above leads us to the interesting technical part of our
discussion, to the analogy with superconductivity and to a comparison
with Umezawa's approach.

\section{The formalism}

The picture of {\it membrane superconductivity}, as opposed to {\it
vortex superconductivity}, can be visualized as ~``islands''~ of
normal vacuum surrounded by a ~``sea''~ of massive $\a(x)$--quanta.
In general, the emergence of different vacuum
phases in the ground state of a physical system is accompanied
by the formation of boundary layers
between the various vacuum domains. In our approach, these
boundaries are approximated by {\it geometrical manifolds} of various
dimensionality (p--branes). This is the point where we depart
from Umezawa's approach: p--branes are introduced at the outset with
their own action functional, and therefore possess their own dynamics
independently of an underlying {\it local} field theory. {\it
Classical} bubble--dynamics has been studied in detail \cite{noi},
and this paper represents a tentative step toward the quantum
formulation. The paradigm of the quantum approach is a {\it line field
theory} introduced several years ago by Marshall and Ramond
as a basis for a second quantized formulation of closed string electrodynamics
\cite{marshram}. We are interested in the case of {\it relativistic,
spatially closed membranes} whose history in spacetime is
represented by infinitely thin (1+2)--dimensional Lorentzian
submanifolds of Minkowski space (M). In a first quantized approach,
membrane coordinates and momenta become operators acting over an appropriate
space of states. However, the non linearity of the theory and the invariance
under reparametrizations
introduce severe problems in the first quantized formulation, e.g.
operator anomalies in the algebra of constraints. At least for closed
membranes, one can bypass these difficulties by considering a {\it
field theory of geometric surfaces} \cite{ho}. If we consider the abstract
space F of all possible bubble configurations, then we are led to consider
a field theory of quantum membranes in which the membrane field is a
reparametrization invariant, complex, {\it functional} of the two--surface $S$
which we assume to be the {\it only} boundary of the membrane
history. Our objective, then, is to introduce and discuss the action
which governs the evolution of quantum membranes regarded as 3--dimensional
timelike submanifolds of Minkowski space. To this end,
our first step is to introduce the {\it 3--volume derivative}
$\delta/\delta \sigma^{\mu\nu\rho}(s)$, which extends the notion of
``loop derivative'' introduced, some years ago, in the framework of the loop
formulation of gauge theories \cite{mig}. The underlying  idea is
this: suppose we attach at a given point $s$ of the surface $S$, an
infinitesimal, closed surface $\delta S$. This procedure is equivalent to a
deformation of the initial shape of $S$ in the neighborhood of $s$, thereby
changing the enclosed volume by an infinitesimal amount $\delta V$.
Then, we define the volume derivative of $\Psi[S]$ through the relation
\beq
\delta\Psi[S]\equiv \Psi[S\oplus \delta S]-\Psi[S]={1\over
3!}\oint_{\delta V}
{\delta\Psi[S]\over \delta\sigma^{\mu\nu\rho}(s)}\,dx^\mu\wedge
dx^\nu\wedge
dx^\rho
\eeq
in the limit of vanishing $\delta V$. This definition is ``local''
to the extent that it involves a single point on the surface. For
the whole $S$ , an averaging procedure is required
\beq
\langle\dots\rangle\equiv \left(\oint_S d^2s \sqrt{\gamma}\right)^{-
1}
\oint_S d^2s \sqrt{\gamma}\left(\dots\right)
\eeq
where, $\gamma=\x\X $ is the determinant of the  metric induced over $S$
by the embedding $ y^\mu=x^\mu(s^i)$ and
$\X=\partial(x^\nu, x^\rho)/\partial(s^1, s^2)$ is the surface tangent
bi--vector. The volume derivative is related to the more familiar
functional variation $\delta/\delta x^\mu(s)$ by the relation
\beq
{\delta\over \delta x^\mu(s)}={1\over 2}\,x^{\nu\rho}
{\delta\over \delta \sigma^{\mu\nu\rho}(s)}.
\eeq
Our second step towards the formulation of the membrane wave
equation, is to introduce the concept of
{\it monodromy} for the $\Psi[S]$ field, since this notion is
directly linked to the physical interpretation of the membrane field.
Our requirement is that $\Psi[S]\equiv A[S]e^{i\Theta[S]}$
be a single valued functional of $S$, i.e. the phase
\beq
\Theta[S]\equiv {1\over 2}\oint_S dx^\mu\wedge dx^\nu
\theta_{\mu\nu}(x)
\eeq
can vary only by $2\pi n$, $n=1,2,...$,
under transport along a ``loop'' in surface space. This condition
constitutes the basis of the analogy with a type--II superconductor.
In order to illustrate the precise meaning of this analogy, it is
convenient to interpret the motion of a bubble in the abstract space
F in which each point corresponds to a possible bubble
configuration. Then, the 3--volume derivative introduced above
represents the spacetime image of the generator of {\it
translations} in F--space and ``classical motion'' in F--space
corresponds to a continuous surface deformation in Minkowski space.
With this understanding, we define a ``line'' in F--space as a one--parameter
family of ``points'', i.e., surface configurations $\{S;t\}$ in physical
space. Let each surface in the family be represented by the embedding equation
$x^\mu=x^\mu(s^1,s^2,t={\rm const.})$, where
$t$ is the real parameter labelling in a one--to--one way
each surface of the family, so that $x^\mu=x^\mu(s^1,s^2;t)$ represents
the embedding of the whole family. However, the same relation
can be interpreted as the embedding of a single three--surface whose $t=$const.
sections reproduce each surface of the family.
In a similar way, we define a ``loop'' of surfaces as a one--parameter
family of surfaces in which the first and the last are identified.
Then, according to our definition of
volume derivative as the spacetime image of the translation
generator in surface space, we define the circulation of $\Theta[S]$ as the
flux of the covariant curl of $\theta_{\mu\nu}(x)$ :
\beqa
\Delta\Theta[S]&\equiv &
{1\over 3!}\oint_\Gamma dx^\mu\wedge dx^\nu \wedge dx^\rho
{\delta\Theta[S]\over \delta\sigma^{\mu\nu\rho}(s)}\nonumber\\
&=&{1\over 3!}\oint_\Gamma dx^\mu\wedge dx^\nu \wedge dx^\rho
\partial_{\,[\mu}\theta_{\nu\rho]}(x) \label{mon1}
\eeqa
where $\Gamma: x^\mu=x^\mu(s^1,s^2;t)$, with
$\partial\Gamma=\emptyset$ represents the spacetime image of the
integration path in surface space.\\
Finally, we define a {\it vortex} line in surface space, as a one--parameter
family of surfaces $\{V;t\}$ for which the amplitude of the membrane field
vanishes, i.e. $\vert\Psi[V_t]\vert=0$. In order to avoid boundary terms
and thus simplify calculations, we assume that the vortex line is closed.
In other words, the spacetime image of the vortex line is a compact three
surface without boundary that we shall denote by $\partial B$. \\
Suppose the test loop of surfaces $\Gamma\equiv \partial\Omega$
surrounds the vortex line $\partial B$, then
the monodromy of $\Psi[S]$ implies the quantization condition :
\beq
\Delta\Theta[S]=
{1\over 3!}
 \oint_\Gamma dx^\mu\wedge dx^\nu \wedge dx^\rho
\partial_{[\,\mu}\theta_{\nu\rho]}=2\pi n\ ,\quad n=1,2,\dots
\label{mon2}
\eeq
If the flux (\ref{mon2}) is quantized, then $\theta$ is a singular function
within $B$. Indeed, using Stokes' theorem, we rewrite (\ref{mon2}) as
\beqa
{1\over 4!}
\int_\Omega dx^\mu\wedge dx^\nu \wedge dx^\rho\wedge dx^\sigma
\partial_{[\,\mu}\partial_\nu\theta_{\rho\sigma]}&=&
{1\over 4!}
\int_B dx^\mu\wedge dx^\nu \wedge dx^\rho\wedge dx^\sigma
\partial_{[\,\mu}\partial_\nu\theta_{\rho\sigma]}\nonumber\\
&=&2\pi n\ne 0\label{sing}.
\eeqa
The  ``Bag'' $B$ is the domain of singularity of the phase 2--form
$\theta_{\mu\nu}(x)$ and represents the spacetime image of the
``vortex interior''. \\

\section {The action}

After the preparatory discussion of the previous section, we assign
to the field $\Psi[S]$ the following action

\beqa
&&S=-{1\over 2\cdot 4!}\int d^4x \F\f+{1\over 3!}\oint [DS]
\langle\vert\D\Psi[S]\vert^2 \rangle\nonumber\\
&&\D_{\mu\nu\rho}\Psi[S]\equiv
\left({\delta\over\delta \sigma^{\mu\nu\rho}(s)}-ig\a\right)
\Psi[S]\label{az}
\eeqa
where, at present, $\oint [DS]\dots$ is a formal way to write the
functional sum over equivalence classes of closed surfaces with
respect to reparametrization invariance, and
$\f=\partial_{[\mu}A_{\nu\rho\sigma]}$ is the gauge field strength
of the rank--three tensor gauge potential $\a(x)$. The shorthand
notation used in (\ref{az}) is convenient but hides some essential
features of the action functional which are worth discussing at this
point. From our vantage point, the key property of the action
(\ref{az}) is its invariance under the extended gauge transformation
\beqa
\Psi'[S]&=&\Psi[S]
\exp\left(i{g\over 2}\oint_S dy^\mu\wedge
dy^\nu\Lambda_{\mu\nu}(y)\right)\\
\a'&=&\a+\partial_{[\,\mu}\Lambda_{\nu\rho]}
\eeqa
This transformation consists of an ``~ordinary~'' gauge term for $\a(x)$,
which is defined over spacetime,
and a non local term for the phase of the membrane functional $\Psi[S]$. \\
The second term in the action contains a spacetime integral which is
not explicitly shown in the expression (\ref{az}). The reason is that a
{\it free} theory of surfaces is invariant under translations of the
{\it center of mass }
\beq
x^\mu=\big{\langle} x^\mu(s)\big{\rangle}
\eeq
so that the membrane action functional in (\ref{az})
 contains the spacetime four--volume as the corresponding zero--mode
contribution. This translational invariance is broken by the coupling
to an ``external'' field $\a(x)$, in which case the four dimensional
zero--mode integral is no longer trivial \cite{ft}. It can be factored out
by inserting the ``~unity operator~'' \cite{kik}
\beq
\int d^4x\,\delta^{4)}\left[{\textstyle\oint_S d^2s\sqrt\gamma
\left(x-x(s)\right)}\right]
\left({\textstyle\oint_S d^2s\sqrt\gamma}\right)=1
\eeq
into the functional integral. Then, we define the sum over surfaces
as a sum over all the surfaces with the center of mass in $x$, and
then we integrate over $x$:
\beqa
\oint D[S](\dots)&=&\int d^4x\oint D[x^\mu(s)]\delta^{4)}
\left[x-\langle x(s)\rangle\right](\dots)\ ,\nonumber\\
&\equiv &\int d^4x\oint D[S_x](\dots)
\eeqa
All of the above applies to the quantum mechanical formulation of
surfaces interpreted as geometric objects. On more
physical grounds, membranes represent energy layers
characterized by a typical thickness, say $1/\Lambda$, which will be
determined later on.
To take into account the finite thickness of a physical
membrane, the singular delta--function which corresponds to the
{}~``thin film approximation''~, has to be smeared into a regular
function sharply peaked around $\langle x^\mu(s)\rangle$. The
simplest representation for such a function is given by a
momentum space gaussian

\beq
 \delta^{4)}\left[x-\langle x(s)\rangle\right]
\rightarrow\delta^{4)}_\Lambda
\left[x-\langle x(s)\rangle\right]\equiv
\int {d^4k\over (2\pi)^4}e^{\disp{-ik_\mu(x^\mu-\langle
x^\mu(s)\rangle)-k^2/2\Lambda}}.
\eeq
However, as long as we work at a distance scale much larger than
the membrane transverse dimension, we can approximate the physical
extended object with a geometrical surface. In what follows we
shall refer to the regularized delta--function only when it is
strictly necessary. With the above
prescriptions, the action (\ref{az})
can be written as the spacetime integral of a lagrangian density
\beq
{\cal S}[\Psi^*,\Psi;\a]=\int d^4x\left\{-{1\over 2\cdot 4!}\f\F
+{1\over 3!}
\oint D[S_x]\langle
\Big\vert {\cal D}\Psi[S]\Big\vert^2\rangle\right\}
\eeq
and the interaction between the membrane field current and the $\a$
potential is described by
\begin{eqnarray}
{\cal S}_{\rm int.}[S;\a]&=&{g\over 2\cdot 3!}\int d^4x\,\oint
D[S_x]\,\left[\,i\langle\Psi^*[S]
{\cal D}_{\mu\nu\rho}^{\!\!\!\!\!\!\!\strut\longleftrightarrow}
\Psi[S]\rangle\right]\A(x)\nonumber\\
&=& {g\over 2\cdot 3!}\int d^4x\oint D[S_x]
\left[\langle i\Psi^*[S]
{{\strut\longleftrightarrow\atop\disp{\delta}}\over\strut
\delta \sigma_{\mu\nu\rho}(s)}\Psi[S]+ g\vert\Psi[S]\vert^2
A^{\mu\nu\rho}\rangle\right]\a(x)\nonumber\\
&\equiv&  {g\over 3!}\int d^4x\, J^{\mu\nu\rho}(x)\a(x)
\label{corr}
\end{eqnarray}
where $g$ is the gauge coupling constant of dimension two in energy units.
As a classical ``charge'', it describes the strength of the
interaction among volume elements of the world--tube swept in spacetime by the
membrane evolution. In our functional field theory $g$ enters as the
interaction constant between the membrane field current and the gauge
potential. Equation (\ref{corr}) exhibits a characteristic {\it London form}
which alerts us about the occurrence of non--trivial vacuum phases.
Indeed, the current implicitly defined in the last step in equation
(\ref{corr}), can be rewritten in the following form

\begin{eqnarray}
J^{\mu\nu\rho}(x)&=&\oint D[S_x]\langle
i\Psi^*[S]{{\strut\longleftrightarrow\atop\disp{\delta}}\over\strut
\delta \sigma_{\mu\nu\rho}(s)}
\Psi[S]\rangle+g\oint D[S_x]\vert \Psi[S]\vert^2
A^{\mu\nu\rho}(x) \nonumber\\
&\equiv& I^{\mu\nu\rho}(x)+{\varphi^2(x)\over g}\A(x)\label{lond}
\end{eqnarray}
where we have introduced the scalar field $\varphi(x)$
\beq
\varphi^2(x)\equiv g^2\oint D[x^\mu(s)]\int {d^4k\over (2\pi)^4}
e^{\disp{-ik_\mu(x^\mu-\langle x^\mu(s)\rangle)-k^2/2\Lambda}}
\vert \Psi[S]\vert^2
\eeq
which we interpret as the {\it order parameter} associated with
membrane condensation in the
same way that the Higgs field is the order parameter associated
 with the boson condensation of point--like objects. In the ordinary
vacuum $\varphi(x)=0$, i.e. there are no centers of mass, and
therefore no membranes. Alternatively, we define a vacuum
characterized by a constant ``density of centers of mass'',
 $\varphi(x)={\rm const.}\ne 0$, as a {\it membrane condensate}.\\
 We will show in Section 4 that the membrane condensate acts as a
{\it superconductor} upon the gauge potential,
turning $\a(x)$ into a massive scalar field. The problem
of surface condensation is thus reduced to studying the distribution
of their representative, pointlike, centers of mass. Conversely, we
show in Section 5 that membrane condensation can be
driven by the $\a(x)$--field quantum corrections alone, and is
accounted for by an effective potential ascribed to the extended
object. This is Umezawa's self--consistency condition transplanted
in our own formalism.

\section{Dynamics of the membrane vacuum and the
formation of bags}

The action (\ref{az}) leads to the pair of coupled field equations

\begin{eqnarray}
&&{\langle}\Big\vert
\left({\delta\over\delta \sigma^{\mu\nu\rho}(s)}-ig\a \right)
\Big\vert^2\Psi[S]{\rangle}=0\label{kg}\\
&&\partial_\lambda F^{\lambda\mu\nu\rho}(x)=g J^{\mu\nu\rho}(x)\
\label{onda}
\end{eqnarray}
which describe the interaction between the membrane field and the
$\a$ gauge potential. Now, we wish to show that the superconducting membrane
condensate contains regions of spacetime, or {\it bags} of non
superconducting vacuum. \\
Recall that in a type--II superconductor the magnetic field is
confined by the superconducting
vacuum pressure within a string--like flux tube. Similarly,
the membrane condensate confines the gauge field strength within
a membrane--like boundary layer surrounding a region of ordinary
vacuum. Indeed, in analogy with the superconducting solution of
scalar QED, we assume the following asymptotic boundary condition for $\Psi[S]$
 \beq
\Psi[S]\equiv {\phi\over g^2}
e^{i\displaystyle{\Theta[S]}}={\phi\over g^2}
\exp\left({i\over 2}\oint_S dx^\mu\wedge
dx^\nu\theta_{\mu\nu}(x)\right)
\eeq
where $\phi$ is a constant.
This is the form of the membrane field when the three volume
enclosed by $S$ is much larger than the three volume of the vortex
spacetime image. Then, from equation (\ref{kg}) we obtain the corresponding
asymptotic form of $\a$:
\beq
\left({\delta\over\delta\sigma^{\mu\nu\rho}(s)}-
g\a\right)\Psi[S]=0\quad
\longrightarrow \a={1\over g}\partial_{[\mu}\theta_{\mu\nu]}.
\label{agauge}
\eeq
Therefore, the flux of $\f$ across a large four dimensional region
$\Omega$ enclosing $B$ is given by

\beqa
q_n&\equiv & {1\over 4!}\int_\Omega \f\,  dx^\lambda\wedge
dx^\mu\wedge dx^\nu \wedge dx^\rho\nonumber\\
&=&{1\over 3!}\oint_\Gamma
 dx^\mu\wedge dx^\nu \wedge dx^\rho \a\nonumber\\
&=&{1\over 3!g}
 \oint_\Gamma dx^\mu\wedge dx^\nu \wedge dx^\rho
\partial_{[\,\mu}\theta_{\nu\rho]}=
{2\pi n\over g}\qquad  n=1,2,\dots \label{flux}
\eeqa
Thus, the physical consequence of the monodromy of $\Psi[S]$
is that the flux of $\f$ through a region enclosing $B$
is quantized in units of $2\pi/g$. \\
In the superconducting phase, equation (\ref{onda}) becomes
\beq
\partial_\lambda F^{\lambda\mu\nu\rho}(x)=-\phi^2\left(
A^{\mu\nu\rho}(x)
-{1\over g}\partial^{[\,\mu}\theta^{\nu\rho]}(x)\right)\equiv -
j^{\mu\nu\rho}(x)
\label{super}
\eeq
in which we have introduced the {\it supercurrent density}
$j^{\mu\nu\rho}(x)$. Equation (\ref{super}) holds only where
$\theta^{\nu\rho}(x)$ is a regular function. In the domain of singularity,
where the partial derivatives of $\theta^{\nu\rho}(x)$ do not commute,
the covariant curl of
$\partial^{[\,\mu}\theta^{\nu\rho]}(x)$ should be interpreted in the
sense of distribution theory. Indeed, if we apply the covariant curl
operator to both sides of equation (\ref{super}), we obtain
\beq
\partial^{[\,\lambda}j^{\mu\nu\rho]}=-
\phi^2\left(F^{\lambda\mu\nu\rho}
-{1\over g}\partial^{[\,\lambda}
\partial^\mu\theta^{\nu\rho]}\right)
\label{curl}
\eeq
The last term in (\ref{curl}) may not be disregarded without violating
(\ref{sing}). Therefore in order to match (\ref{flux}) with (\ref{sing}),
we define
\beqa
\partial^{[\,\lambda} \partial^\mu\theta^{\nu\rho]}(x)& \equiv&
{q_n g\over\Omega_B}\epsilon^{\lambda\mu\nu\rho}
\int_B d^4\xi\,\delta^{4)}\left[x-z(\xi)\right]\nonumber\\
&\equiv& {q_n g\over\Omega_B} J_B^{\lambda\mu\nu\rho}(x)
\label{bagcurr}
\eeqa
where, $\Omega_B$ and $J_B^{\lambda\mu\nu\rho}(x)$
are, respectively, the bag four--volume and the bag current.
Thus, the supercurrent can be determined from the equation
\beq
\partial_\lambda\partial^{[\,\lambda}j^{\mu\nu\rho]}=
-\phi^2\left(j^{\mu\nu\rho}
-{q_n\over \Omega_B}\partial_\lambda J_B^{\lambda\mu\nu\rho}\right)
\eeq
by means of the Green function method:
\beq
j^{\mu\nu\rho}(x)=-{\phi^2 q_n\over\Omega_B}
\epsilon^{\mu\nu\rho\sigma}\int_B d^4z\,
\partial_\sigma\, G(x-z; \phi^2)
\label{superc}
\eeq
where $G(x-z)$ is the scalar Green function
\beq
\left[\partial^2 +\phi^2\right]G(x-z;\phi^2)=\delta^{4)}(x-z).
\eeq
Then, from equation (\ref{onda}), (\ref{curl}) and
(\ref{superc}), we find the form of the confined gauge field
\beq
F^{\lambda\mu\nu\rho}(x)=-{\phi^2
q_n\over\Omega_B}\epsilon^{\mu\nu\rho\sigma}
\int_B d^4z\, G(x-z; \phi^2).
\label{conf}
\eeq
The analogy between the membrane vacuum and a type--II
superconductor now seems manifest: in the ordinary vacuum $\a$ does
not propagate any degree of freedom.
Rather, it corresponds to a uniform energy background. However, in
the superconducting phase $\a$ becomes a dynamical field describing
a massive, spin--0 particle \cite{aa}. The source for the massive
field is the bag current (\ref{bagcurr}). In a boson particle
condensate, the magnetic
field is confined to a thin flux tube surrounding the vortex line;
in the membrane condensate, the $\f$--field is confined within the membrane
which encloses the ordinary vacuum bag. The gauge field
provides the ``skin'' of the bag. To complete the analogy, in the
next section we show that the thickness
 of the membrane is given by the inverse of the
dynamically generated mass of $\f$.\\

\section{The Coleman--Weinberg mechanism of mass generation}

The scenario emerging from the last section is based on the assumption that the
$\varphi(x)$--field can acquire a non vanishing vacuum expectation value.
Note that there is no potential term for $\varphi(x)$ in the
classical action (\ref{az}). However, we wish to show that a self--consistent
potential may originate from the quantum fluctuations of the $\a(x)$
field leading to a non vanishing vacuum
expectation value for the order parameter. On the technical side, this means
to compute the one--loop effective potential for the $\varphi(x)$ field
by integrating $\a$ out of the functional integral
\beq
Z=\int D \left[\Psi^*[S]\right]D\left[\Psi[S]\right][D\a(x)]
\exp\left(i{\cal S}/\hbar\right).
\eeq
The quantization of higher rank gauge fields is a lengthy procedure
involving a sequence of gauge fixing conditions together with
various generations of ghosts \cite{town}. In principle, these terms
should be included in the functional measure in the action
functional. However, in our case they are unnecessary since we know
already that in the superconducting phase $\a$ describes
a massive scalar degree of freedom which is the {\it only physical degree of
freedom}. Thus, the effective potential is
\beqa
\veff&=&{1\over 2}{\rm Tr}\,\ln\left[\left(\partial^2+\phi^2\right)/
\Lambda^2\right]+{\delta\rho(\Lambda)\over 2}\phi^2+
{\delta\lambda(\Lambda)\over 4!}\phi^4\nonumber\\
&=&{1\over 32\pi^2}\left[\phi^2\Lambda^2+{\phi^4\over 2}\left(
\ln\left(\phi^2\over\Lambda^2\right)-{1\over 2}\right)\right]+
{\delta\rho(\Lambda)\over 2}\phi^2+{\delta\lambda(\Lambda)\over
4!}\phi^4\nonumber\label{peff}.
\eeqa
The ultraviolet divergences of the one--loop determinant have been
regularized through the cutoff $\Lambda$, and the two counterterms
$\delta\rho(\Lambda)$ and $\delta\lambda(\Lambda)$ are fixed by the
renormalization conditions
\beqa
&&\left({\partial^2\veff\over\partial\phi^2}\right)_{\phi=0}=0\label
{v2}\\
&&\left({\partial^4\veff\over\partial\phi^4}\right)_{\phi=g/\mu}=0
\label{v4}
\eeqa
in which $\mu$ appears as an arbitrary renormalization scale. The scalar field
$\varphi(x)$ has no classical dynamics of its own, i.e., it possesses
no kinetic or potential term.
This is the reason for imposing the two conditions (\ref{v2}),(\ref{v4}):
equation (\ref{v2}) is the characteristic Coleman--Weinberg condition
\cite{cw} ensuring that the mass of the gauge field is non vanishing only
in the condensed phase $\phi\ne 0$; equation (\ref{v4}) follows from the
absence of a classical quartic self--interaction. Of course, the physical
properties of the system are insensitive to the choice of the
renormalization condition. Then, with our choice,
we find the {\it Coleman--Weinberg potential for membranes}
\beq
\vcw={\phi^4\over 64\pi^2}\left[\ln{\phi^2\mu^2\over g^2}
-{25\over 6}\right].
\eeq
The absolute minimum of $\vcw$ corresponds to a super--conducting phase
characterized by a vacuum expectation value of the order parameter
\beq
\langle\phi\rangle^2={g^2\over\mu^2} e^{\disp{{11\over 3}}}
\label{min}
\eeq
and by a dynamical surface tension
\beq
 {\rho_R^2\over g^2}\equiv
\left({\partial^2\vcw\over\partial\phi^2}\right)_{\phi=\langle\phi\rangle}=
{g^2\over 8\pi^2\mu^2} e^{\disp{{11\over 3}}}.
\eeq
The factor $\langle\phi\rangle^2$ as given by (\ref{min}) is also
the square of the dynamically generated mass for $\a(x)$.
Hence, the physical thickness of the membrane, is of the order of
$\langle\phi\rangle^{-1}$, and the dynamically generated surface tension is
$\rho_R=g\langle\phi\rangle/\sqrt{8\pi^2}$. This quantity is
positive, so that the bubbles of ordinary vacuum tend to collapse in
the absence of a balancing internal pressure. In the case of
{}~``hadronic bags''~, this internal pressure is provided by the
quark--gluon complex. In any event, the picture
of the superconducting membrane vacuum is
strongly reminiscent of the {\it classical} dynamics of a closed membrane
coupled to its gauge partner i.e., $\a(x)$ \cite{noi}. In both cases,  vacuum
bubbles created in one vacuum phase evolve and die in a different
vacuum background. This suggests a new possibility of quantum vacuum
polarization via the creation and annihilation of whole domains of
spacetime in which the energy density is different from that of the
ambient spacetime. As a matter of fact, the
novelty of our field model is the onset of a new type of ``Higgs
mechanism for membranes'' triggered solely by quantum fluctuations.
The effect of such fluctuations can be accounted for by an effective
potential. As in Umezawa's approach, this effective potential is
consistent with the dynamical generation of a bag with surface
tension out of the vacuum.

\end{document}